\documentclass[twocolumn,nofootinbib,superscriptaddress]{revtex4-1}

\pdfoutput=1

\usepackage{slashed}

\usepackage[normalem]{ulem}

\usepackage{color}

\usepackage{epsfig}
\usepackage{epstopdf}
\usepackage{epsf}
\input{epsf.sty}
\usepackage{graphicx,amsmath,amssymb}
\usepackage{epsfig}
\allowdisplaybreaks

\def\ei{\end{itemize}}
\def\be{\begin{equation}}
\def\ee{\end{equation}}
\newcommand{\bea}{\begin{eqnarray}}
\newcommand{\eea}{\end{eqnarray}}

\usepackage{bbm,bm,amsmath,amssymb}

\usepackage{hyperref}

\newcommand{\rf}[1]{(\ref{#1})}
\newcommand{\la}{\lambda}

\usepackage{float}

\begin{document}

\title{ Gaugino Condensation and Geometry of the Perfect Square }
\author{Renata Kallosh}
\email{kallosh@stanford.edu}
\affiliation{Stanford Institute for Theoretical Physics and Department of Physics, Stanford University, Stanford,
CA 94305, USA}

 \begin{abstract}
Gaugino condensation plays a crucial role in the formation of dS vacua. However, the theory of this effect is still incomplete. In 4d the perfect square nature of gaugino couplings follows from  the square of auxiliaries in the supergravity action.
We explain here why   the supersymmetric non-Abelian Dp-brane action,  which is the basis for the theory of  10d gaugino condensation, must have a 4-gaugino coupling. This term in the Einstein-Yang-Mills 10d supergravity  is a part of the perfect square,  mixing a 3-form with gaugino bilinear.
The perfect square term follows from the  superspace geometry, being a square of the  superspace torsion. The supercovariant equation of motion for gaugino on the non-Abelian Dp-brane also involves a supertorsion in agreement with the perfect square term in the action.

 \end{abstract}

\maketitle


\section{Introduction}

The issue of supersymmetry in Dp-brane actions is well understood for a single Dp-brane. The corresponding action  constructed in \cite{Cederwall:1996pv,Cederwall:1996ri,Bergshoeff:1996tu} has a local fermionic  $\kappa$-symmetry.   When this local   $\kappa$-symmetry  is  gauge-fixed as in \cite{Aganagic:1996nn, Bergshoeff:1997kr},  one recovers the Abelian supersymmetric Dp-brane action, which breaks the maximal supersymmetry of type II supergravity,  half of it being realized linearly. 

In \cite{Marolf:2003ye,Camara:2004jj,Martucci:2005rb,Baumann:2010sx,Dymarsky:2010mf}  the analysis of fermions on D7 brane proceeds starting with Abelian supersymmetric brane, where the coupling of the background 3-form field to the bilinear of the fermions was established. The 4-fermion non-derivative coupling is absent  in the Abelian Dp-brane action,  
as one can see by a direct inspection of the gauge-fixed $\kappa$-symmetric brane action  \cite{Cederwall:1996pv,Cederwall:1996ri,Bergshoeff:1996tu,Aganagic:1996nn, Bergshoeff:1997kr}.

The concept of gaugino condensation in 10d is based on properties of N coincident D7-branes. The problem of finding the 
supersymmetric  non-Abelian action for the N coincident Dp-branes is not fully solved yet. In particular, the 4-fermion non-derivative coupling was  not investigated  in  \cite{Marolf:2003ye,Camara:2004jj,Martucci:2005rb,Baumann:2010sx,Dymarsky:2010mf} in this context.

This issue attracted some attention after  the recent 10d investigation of the KKLT mechanism in  \cite{Moritz:2017xto} (see also \cite{Gautason:2018gln}). This investigation was based on a number of assumptions debated in \cite{Cicoli:2018kdo,Akrami:2018ylq,Kallosh:2018wme,Kallosh:2018psh},  and the final   conclusions of  \cite{Moritz:2017xto,Gautason:2018gln} turned out to be inconsistent with the results of the 4d KKLT analysis  \cite{Kallosh:2018wme,Kallosh:2018psh,Kallosh:2019axr}.

One of the  assumptions made in  \cite{Moritz:2017xto,Gautason:2018gln} was the absence of the 4-fermion non-derivative coupling on D7 branes. This assumption was recently challenged in \cite{Hamada:2018qef}. The proposal  how to incorporate such terms, which was made in  \cite{Hamada:2018qef}, was motivated by the heterotic string theory and Horava-Witten model \cite{Horava:1995qa,Horava:1996ma,Horava:1996vs}, which both have the corresponding 4-gaugino couplings associated with the perfect square structure of Einstein-Yang-Mills supergravity in 10d. 
 
Here we will explain why  the non-Abelian generalization of Dp-brane action  must have a 4-fermion non-derivative coupling. We will also show that the supersymmetric equation of motion for gaugino has a cubic gaugino coupling, which vanishes only in the Abelian case of a single Dp-brane. The perfect square structure underlying these features of the Dp-branes is a consequence of the superspace geometry.
These results are quite general, they apply to   type IIB theory and to the KKLT scenario.  
They support and further develop the proposal made in  \cite{Hamada:2018qef}.

\section{Gaugino's in  D${\rm \bf p}$-brane actions} \label{10d}

The Abelian Dp-brane action upon gauge-fixing a local $\kappa$-symmetry has an unbroken  supersymmetry. Consider the   coupling of the background 3-form with the spinor field on the brane  in the form given in \cite{Martucci:2005rb}, where the gauge-fixed action for the Dp-brane is given in approximation quadratic in fermions
\be
\int  d^{p+1}  x  \,    G_{\mu\nu\rho} \, \bar \la \Gamma^{\mu\nu\rho}  \la \ .
\label{coupling}\ee
Looking at the complete  gauge-fixed Dp-brane action \cite{Aganagic:1996nn, Bergshoeff:1997kr} one finds that there are no 4-fermion terms without derivatives on supersymmetric Dp-brane worldvolume.  But there is also no disagreement with a possibility to add to the action these terms in the form $(\bar \la \Gamma^{\mu\nu\rho}  \la)^2 $ since they vanish  according to  Fierz identities,  as shown in \cite{Nilsson:1981bn,Bergshoeff:1981um},
\be
(\bar \la \Gamma^{\mu\nu\rho}  \la)^2 =0 \ .
\label{Fierz}\ee
The situation changes dramatically when one is trying to figure out the form of the supersymmetric non-Abelian Dp-brane action. The coupling of a background 3-form \rf{coupling} to gaugino's generalizes to 
\be
\int  d^{p+1}  x  \,    G_{\mu\nu\rho} \,  \mathrm{Tr}\, \bar \la \Gamma^{\mu\nu\rho}  \la \ .
\label{couplingN}\ee
In the non-Abelian case, the Fierz identity shows that the 4-fermion term is not vanishing anymore,
\be ( \mathrm{Tr}\, \bar \la \Gamma^{\mu\nu\rho}  \la)^2 \neq 0 \ .
\label{FierzN}\ee
What is the generalization of the action \rf{coupling}   to the non-Abelian case?  The answer is defined by a supersymmetry of the effective low-energy action of supergravity with Dp-brane sources. We will show below that   the 4-fermion terms with non-derivative coupling shown in eq. \rf{FierzN} must be present.

When a Dp-brane is described as a local source to the bulk supergravity action the issue of $\delta ^{9-p} (0)$ divergences is important, and has to be dealt with, as explained in  \cite{Hamada:2018qef} in the case of D7 branes with $\delta ^{2} (0)$ divergences. Following \cite{Horava:1995qa,Horava:1996ma,Horava:1996vs}, they argue that  the infinite terms cancel due to presence of 4-fermion terms, whereas the finite terms also involve  bilinear  as well as  4-fermion gaugino coupling.

   We will start with the space filling  coincident  D9 branes where the issue of  the existence/non-existence of the 4-fermion gaugino term on the Dp-brane worldvolume is disentangled from the issue of divergences.
   
   \section{The 4-fermion interaction in 4${\rm \bf d}$ Einstein-Yang-Mills supergravity}
   In general 4d supergravity the role of gaugino condensates with regard to spontaneous breaking of local supersymmetry was explained in \cite{Ferrara:1982qs} based on the properties of the auxiliary fields $F^\alpha$ of the chiral multiplets discovered in \cite{Cremmer:1982en}. The supergravity action is quadratic in auxiliary fields
 \be
 {\cal L}_{aux}= - F^\alpha g_{\alpha\bar \beta} \bar F^{\bar \beta} \equiv |F|^2\, . 
\label{aux} \ee  
Here the  on-shell value of  auxiliary field is
   \be
F^\alpha = - e^{K/2} g^{\alpha \beta} \overline \nabla_{\bar \beta} \overline W     + {1\over 4} \bar f_{AB\bar \beta} g^{\bar \beta \alpha} \bar \lambda^A P_L   \lambda^B
\label{aux1}   \ee
 as one can see in eq. (3.14) and in eq. (4.4) in  \cite{Cremmer:1982en},  but employ 
 notation of \cite{Freedman:2012zz}. As in eq. (3) in \cite{Ferrara:1982qs} we omit the fermions from the chiral multiplets here and only keep fermions from  the Yang-Mills multiplets.
 
 Eqs. \rf{aux}, \rf{aux1}  give the first indication of the perfect square nature of the 2-fermion and 4-fermion coupling in the action.  
This phenomenon is very general\footnote{I am grateful to S. Ferrara for explaining this to me.} and underlines the related examples in 10d theory and in supersymmetry breaking by Dp-branes which we will discuss below.
          
\section{ The 4-fermion interaction in 10${\rm \bf d}$ Einstein-Yang-Mills supergravity}
Once the Dp-brane is added as a local source to the 10d supergravity action, half of supersymmetry is broken. An effective supergravity action describing this situation in an example of a single D9-brane, is the action of the 10d  Einstein-Maxwell supergravity \cite{Bergshoeff:1981um}. The gravitational multiplet  includes the graviton, gravitino, dilatino, a 2-form field, and a dilaton: $e_\mu^m, \psi_\mu, \chi, A_{\mu\nu}, \phi$. A matter Maxwell multiplet includes an Abelian vector and a gaugino: $A_\mu, \la$. 

The total Einstein-Maxwell supergravity action with ${\cal N}=1$ 10d supersymmetry is \cite{Bergshoeff:1981um}
\bea\label{EMaction}
 e^{-1} {\cal L}^{\rm EM} = &&- {1\over 2}  R - {3\over 4}\phi^{-{3\over 2}} \Big ( F_{\mu\nu\rho}^M      - {\sqrt 2\over 24}  \, \phi^{{3\over 4}} \, \bar \la \Gamma_{\mu\nu\rho} \la\Big )^2 \cr
 \cr
&& - {9\over 16} \Big ({\partial_\mu \phi\over \phi}\Big )^2 -{1\over 4} F_{\mu\nu}^2 -{1\over 2} \bar \la \gamma^\mu D_\mu \la  + \dots
\eea
Here terms with $\dots$ depend on gravitino $\psi_\mu$ and on a dilatino $\chi$. The 3-form field is
$
F_{\mu\nu\rho}^M= \partial_{[\mu} A_{\nu\rho]} -{1\over \sqrt 2} A_{[\mu} F_{\nu\rho]} \ ,
$
and $F_{\mu\nu}= \partial_{\mu} A_{\nu} - \partial_{\nu} A_{\mu} $.
The 4-fermion term here is vanishing in Abelian case of just one matter multiplet \rf{Fierz}. Therefore the perfect square term in \rf{EMaction} is optional, one could have had it in the form without a 4-fermion coupling, in which case it would not look like a perfect square term. This is exactly  the case in the Einstein-Maxwell supergravity \cite{Bergshoeff:1981um}, where instead of  the first line of the action 
\rf{EMaction} one has
\bea\label{EMaction1}
- {1\over 2}  R - {3\over 4}\phi^{-{3\over 2}} \Big ( (F_{\mu\nu\rho}^M)^2      - {\sqrt 2\over 12} \, \phi^{{3\over 4}} \,  F^{\mu\nu\rho M} \bar \la \Gamma_{\mu\nu\rho} \la\Big ) 
\eea
and the 4-fermion gaugino term is absent.

The D9 brane action at the level where we neglect higher derivative non-linear terms of the Born-Infeld nature is precisely the supersymmetric Maxwell action in the background of ${\cal N}=1$ supergravity. Therefore
the supersymmetric action \rf{EMaction} can be viewed as an effective low energy approximation of the single  D9-brane interacting with gravity, which has ${\cal N}=1$ 10d  supersymmetry. 

If, on the contrary,  we would like to have a whole non-linear Dp-brane action in the background of type IIB supergravity, we would be able to start with a  $\kappa$-symmetric action \cite{Cederwall:1996pv,Cederwall:1996ri,Bergshoeff:1996tu}, gauge-fix local fermionic $\kappa$-symmetry \cite{Aganagic:1996nn, Bergshoeff:1997kr}, and we would also be able to identify the fermionic terms on the Dp-brane. For a single Dp-brane the 4-fermion terms are absent, in agreement with \rf{Fierz}. Meanwhile,  a complete $\kappa$-symmetric action for coinciding multiple  Dp-branes, which are supposed to describe a non-perturbative gaugino condensation, is not known.

Therefore we can look at the gaugino condensation problem from the perspective of the Einstein-Yang-Mills 10d supergravity \cite{Chapline:1982ww}, but correct the omission of the 4-gaugino term there (we skip terms depending on gravitino and dilatino):
\bea\label{EYMaction}
&e^{-1} {\cal L}^{\rm EYM} = - {1\over 2}  R - {3\over 4}\phi^{-{3\over 2}} \Big ( F_{\mu\nu\rho}^{YM}   - {\sqrt 2\over 24} \, \phi^{{3\over 4}} \, \mathrm{Tr} \,  \bar \la \Gamma_{\mu\nu\rho} \la\Big )^2 \cr
 \cr
&- {9\over 16} \Big ({\partial_\mu \phi\over \phi}\Big )^2 -{1\over 4}  \mathrm{Tr} \, (F_{\mu\nu}^{YM})^2 -{1\over 2} \mathrm{Tr} \,  \bar \la \gamma^\mu D_\mu \la \, .
\eea
 The 3-form field is now
$
F_{\mu\nu\rho}^{\rm YM} = \partial_{[\mu} A_{\nu\rho]} -{1\over \sqrt 2} \mathrm{Tr} \Big ( A_{[\mu} F_{\nu\rho]} - {2\over 3} A_{[\mu} A_{\nu}  A_{\rho]} \Big ) 
$
and $F_{\mu\nu}^{YM}= \partial_{\mu} A_{\nu} - \partial_{\nu} A_{\mu} + [ A_{\mu},  A_{\nu}]$.
The 4-fermion term here is not vanishing in the non-Abelian case, see \rf{FierzN}.

In   \cite{Chapline:1982ww} the 4-fermion gaugino term was omitted even in the non-Abelian case,  which was corrected in \cite{Dine:1985rz}.   The emphasis in \cite{Dine:1985rz} was on the fact that the corresponding 4-fermion term originates from a heterotic string theory where the computation of the low energy scattering of gaugino's has revealed its presence.

\section {The geometric superspace origin of the perfect square}
One could perform a brute force computation of the local supersymmetry of the Chapline-Mantone action \cite{Chapline:1982ww} (or one can compute the 4-fermion amplitude in the heterotic string) and find out that the 4-fermion gaugino term is required. For example, one can show that the terms with two gaugino's and a gravitino and a dilatino, $\la^2 \psi \chi$ and the terms with two gaugini's and two dilatino's, $\la^2 \chi^2$, in the non-Abelian case have a variation of the form
\be
\delta {\cal L}^{M} \sim  \mathrm{Tr} \bar \la \gamma^{abc} \la \, \bar \chi X^{abc a'b'c'} \epsilon \,   \mathrm{Tr}\bar \la \gamma^{a'b'c'}\la \ ,
\label{M}\ee
where $X^{abc a'b'c'} $ is a complicated combination of the $\gamma$ matrices. In Abelian case a cubic in $\la$ expression  $\bar \la \gamma^{abc} \la \, \, \gamma_{ab} \la$ vanishes and it can be used to show that $ \delta {\cal L}^{M}=0$ in $\la^4 \chi \epsilon$ sector. However, in the non-Abelian case the expression in \rf{M} does not vanish, and the cooperation with the variation from the 4-fermion term $( \mathrm{Tr} \bar \la \gamma^{abc} \la)^2$, with a specific factor in front, is required.

This can be confirmed by a laborious work, and one  would also notice the following. Once the  $( \mathrm{Tr} \,  \la \, \la )^2$ term with the correct coefficient is added to the action, the 3-form and bilinear gaugino form a perfect square as shown in eq. \rf{EYMaction}. In this procedure the appearance of the perfect square in the action is not clear, it looks like an accident.

However, there is a geometric reason for the  perfect square in Einstein-Yang-Mills 10d supergravity in the context of the superspace geometry: the perfect square in the action is a square of the superspace torsion tensor at $\theta=0$, where $\theta$ is the Grassmann coordinate of the superspace.

The 10d superspace 
was constructed in \cite{Kallosh:1985cd,Nilsson:1985si,Nilsson:1986md}. We represent it here in slightly more convenient notation.  
One starts with a superspace with coordinates $z^M= (x^\mu, \theta^m)$. In this superspace we introduce the vielbein $E^A$,   the spin connection $\omega_A{}^B$, the two-form potential $B$, and the Yang-Mills one-form $A$. The tangent space indices $A, B$ split into bosonic  indices $a,b$  and fermionic   indices $\alpha, \beta$. We define an exterior derivative covariant with respect to general superspace coordinate transformations,  as well as local Lorentz and local Yang-Mills transformations: $D= d+\omega +A $.  The standard definition of the torsion and curvature in the superspace follows, as well as a three-form and a Yang-Mills two-form, see eqs. (3-6) in \cite{Kallosh:1985cd} for details. The torsion $T^A$, the curvature $R_A{}^B$,  the supergravity three-form $H$ and the Yang-Mills two-form $F$ satisfy the superspace Bianchi identities (BI)
\bea
&D T^A = - E^B \wedge R_B{}^A\, , \quad DR_A{}^B=0\, , \cr
\cr
 &DH = \mathrm{Tr} \, F\wedge F \, , \quad DF=0 \ .
\eea
For example the torsion two-form defines the torsion tensor
\be
T^A = {1\over 2} E^B \wedge E^C \, T^A_{CB} \ .
\ee
To find the solutions of the BI one has to impose certain constraints on the coupled Einstein-Yang-Mills superspace, for example that $T_{\alpha \beta}^a = -i \gamma^a_{\alpha \beta}$ etc, as shown in Table 1 in  \cite{Kallosh:1985cd}. The solution of BI was presented in 
 \cite{Nilsson:1985si,Nilsson:1986md} and is given in terms of the following superfields. There is a superfield with the first component, which is a dilaton, $\varphi (x)$,  and all higher components are computed by the repeated action of spinorial derivatives $D_\alpha$. There is a YM superfield with the first component being a gaugino field $\la(x)$. Finally, one more superfield is required with the first component being a combination of the three-form and a gaugino bilinear
 \be
 Y_{abc} \equiv  -{1\over 72} H_{abc} - {2\over 3} \Lambda_{abc}\ ,
\label{Y} \ee
where 
\be \mathrm{Tr} \,   \la_\alpha \la_\beta \equiv  \gamma^{abc}_{\alpha \beta} \Lambda_{abc} \ .
\ee 
The component of the superspace torsion $T_{a\beta}{}^\gamma (x,\theta)$  which solves the superspace BI depends both on the dilaton superfield $\varphi(x,\theta) $ as well as on the superfield $ Y_{abc} (x,\theta) $,
 \be
T_{a\beta}{}^\gamma (x,\theta) = -  {1\over 2} (\gamma^b \gamma_a)_\beta{}^\gamma D_b  \varphi  - \Gamma_a{}^{bcd} Y_{bcd} \ .
\label{torsion} \ee
There is also another third rank antisymmetric tensor superfield,  $ K_{abc} = -{1\over 72} H_{abc} - {i\over 3} \Lambda_{abc} -{5i \over 2} \chi_{abc}$, which includes also a dilatino bilinear $\chi_{abc}$ and has a   mix of $H_{abc}$ with  $ \Lambda_{abc}$, which is different from the one in  $Y_{abc}$. It is a
 second spinorial derivative of the dilaton superfield, and it does not have a superspace tensor properties like  $Y_{abc}$, which represents the superspace torsion.
 
 The component action perfect square term in \rf{EYMaction}
originates from the square of the superspace torsion term
\be
 {\cal L}^{\rm EYM}  \supset T_{a\beta}{}^\gamma   T_\gamma{} ^{a\beta}(x, \theta) |_{\theta=0}  \sim  Y_{abc}  Y^{abc}(x)  \ ,
 \label{T2}
 \ee
 where we have neglected  the terms with derivatives on a dilaton which are present in a superspace torsion \rf{torsion}.

 The Einstein equation for the 10d curvature also depends only on the same perfect square
 \be
 R_{10} \supset Y_{abc}  Y^{abc}(x) +\cdots
 \ee
 
 Thus the superspace geometry clearly explains that  supersymmetry does require the 4-fermion term $( \mathrm{Tr}\, \bar \la \Gamma^{\mu\nu\rho}  \la)^2$ in the action, and moreover, it has to come with the coefficient which forms a perfect square in the action, as shown in eqs. \rf{EYMaction}, \rf{T2}. Therefore, the solution with gaugino condensation, which makes the first component of the superfield vanishing,
 \be
  Y_{abc} = - {1\over 72} H_{abc} - {2\over 3}\langle  \Lambda_{abc} \rangle =0 \, , 
\label{Y0} \ee
 is consistent with local supersymmetry of the theory. In such case, the term in the action $ Y_{abc}  Y^{abc}$ does not contribute to the vacuum energy.

 Thus in the case of  a non-compact geometry, described by the superspace in  \cite{Kallosh:1985cd,Nilsson:1985si,Nilsson:1986md}, gaugino condensation requires a non-vanishing 3-form. It backreacts so that the total square vanishes. This is in agreement with the earlier observations on this in \cite{Dine:1985rz}.

We conclude that  if a complete supersymmetric non-linear non-Abelian Dp-brane action is available, 
it must have on its world volume the 4-fermion term to provide 
the perfect square term shown in eq. \rf{EYMaction} in the D9 case. This is necessary to satisfy a condition of supersymmetry of a non-Abelian vector supermultiplet interacting with supergravity. 

\section{Gaugino Equation of Motion on D${\rm \bf p}$-brane}
  For D7  brane, it is more difficult  to use the perfect square structure of the action in the component form  in \rf{EYMaction} or in the superspace form in  \rf{T2}, as we did it for the D9 brane. Indeed, the 10d action originates from the supergravity action $\int d^{10} x\, G_{abc} G^{abc} $, whereas the gaugino dependent terms originate from the brane action $\int d^{8} x\,  G_{abc} \mathrm{Tr}  \bar \la \gamma ^{abc} \la$ and $\int d^{8} x\,  (\mathrm{Tr}  \bar \la \gamma ^{abc} \la)^2$. Therefore we have to deal with some singular terms from the local sources.

Of course, one can rely on the simple fact that the 4-fermion term existing on D9 should remain on D7 after dimensional reduction.
 To support this argument   that the term $\int d^{8} x\,  (\mathrm{Tr}  \bar \la \gamma ^{abc} \la)^2$ must be present in D7 brane action, we may look at the supersymmetric equation of motion for gaugino which  interacts with the supergravity background. Such an equation was derived  in the context of the superspace geometry  \cite{Kallosh:1985cd,Nilsson:1985si,Nilsson:1986md} in the non-compact 10d space.
 
 The manifestly supersymmetric field equation for  gaugino is presented in eq. (21) in \cite{Nilsson:1986md}. It  depends on the background 3-form only via the torsion tensor $Y_{bcd}$ 
 \be
 \gamma^a \hat D_a \la= 3  Y_{abc} \gamma^{abc} \la =   3 \large(-{1\over 72} H_{abc} - {2\over 3}  \Lambda_{abc} \large)   \gamma^{abc} \la \, .
\label{EOM} \ee
 Here $\hat D_a= D_a - 3 D_a \phi$. The 3-gaugino coupling term in the manifestly supersymmetric field equation  for  gaugino in \rf{EOM}  is given by
 \be
 \gamma^{abc} \Lambda_{abc}   \la
 \sim \mathrm{Tr}  (\bar \la \gamma_{abc}  \la)  \gamma^{abc} \la\ ,
\label{cubic}  \ee
since $\Lambda_{abc}\sim \mathrm{Tr}  \bar \la \gamma_{abc} \la$.
This cubic gaugino coupling  vanishes in the Abelian case  where $  (\bar \la \gamma_{abc}  \la)  \gamma^{ab} \la=0$, but in the non-Abelian case the cubic coupling is not vanishing. This confirms unambiguously the presence of the 4-gaugino coupling on a non-Abelian Dp-brane.

\section{String theory and Finite Volume Issues}

The next conceptual step away from supergravity, defined in the infinite 10d space-time, was an observation in  \cite{Horava:1995qa,Horava:1996ma,Horava:1996vs} about an M-theory compactified on a one-dimensional finite size interval. The localized sorce  in this case of the form $\sim \delta (x^{11})$ is present, however it is argued in  \cite{Horava:1995qa,Horava:1996ma,Horava:1996vs} that supersymmetry requires the presence of a quartic gaugino term in the action. In  \cite{Horava:1996vs} it was suggested to  shift the field strength four-form $G_{IJK11}\equiv G_{abc}$  by a term supported at the boundary and bilinear in the gauginos, and to define the modified 3-form field as follows
 \be
 \tilde G_{abc} = G_{abc} + c\,  \delta (x_{11}) \, \Lambda_{abc} \ .
  \ee
The effective action depends  on the 3-form $G_{abc}$ only via the perfect square,
\be
{\cal L}  \supset ( \tilde G_{abc})^2 \ .
\ee
The  presence of this 4-fermion term in the action  precisely cancels the divergence due to a perfect square structure analogous to the one we explained above. In a sense, the 10d non-Abelian supersymmetry is an underlying reason for the 4-fermion term in the action.

The 4-fermion term plays an important role in gaugino condensation, as shown in \cite{Hamada:2018qef}. 
They have proposed that gaugino condensation in D7-branes should be described by the effective action in their eqs. (39)-(41).  The presence of the 4-gaugino coupling on D7 brane, according to   \cite{Hamada:2018qef}, resolves the issues of divergences and allows a comparison with the 4d theory.

In view of the compact volume of compactification, things are more complicated than in the case of the infinite 6d space described above in our eqs. \rf{EYMaction} and \rf{T2} for  10d 
Einstein-Yang-Mills supergravity, and for non-Abelian D9 branes. 

However, the  perfect square structure of the effective action in non-Abelian D7 is not just reminiscent of heterotic strings or M-theory compactified on the interval,  the perfect square terms in the action in all of these cases  are actually {\it inherited from the 10d Einstein-Yang-Mills superspace geometry}. 

If the D9-branes are compactified on  a torus of a square area $A_T^2= L^2$, and the total internal space is $X= \Sigma \times T^2$ with $V_\Sigma$ being the finite volume of the $\Sigma$-manifold, one would expect that our action \rf{EYMaction} would lead to eqs. (39)-(41) in  \cite{Hamada:2018qef}. This would result in cancellation of the relevant infinities and in clarification of the gaugino condensation on D7-branes.

Specifically, once the conjectured supersymmetric  action of D7 non-Abelian branes is derived from a compactification of D9 branes on $T^2$, it is plausible that the remaining finite terms in the action are of the form given in eq. (41) in  \cite{Hamada:2018qef},
\be
- g_s \Big | G_3^{(0)} + {\lambda \lambda \over \sqrt{g_s} A_{T^2} }\overline \Omega_3 \Big |^2 \ .
\label{AG}\ee
If one admits the terms bilinear in spinors on a non-Abelian D7 brane, also the quartic terms must be present, as   follows from \rf{EYMaction} and from \rf{EOM}, \rf{cubic}. 

The technical issues with singularities  may still require some clarification. The issue of supersymmetry in singular spaces is complicated,  as was shown in the past in a different context in \cite{Bergshoeff:2000zn}.

\section{Relation to 4${\rm \bf d}$ supersymmetry}
 It is  known that type IIB supergravity with local  sources, which involve calibrated Dp-branes and O-planes, leads to a 4d 
${\cal N}=1$ supergravity. String theory constructions in 10d space compactified on calibrated manifolds were mostly compared with 4d ${\cal N}=1$ supergravity with chiral multiplets. The  dictionary between the Kahler potential $K(z, \bar z)$ and the  holomorphic superpotential $W(z)$ and   10d theory  was established. Moreover, in presence of pseudo-calibrated anti-Dp-branes, the 4d supergravity also involves a nilpotent chiral superfield, see \cite{Kallosh:2018nrk} and references therein. Meanwhile, the presence of the vector multiplets in 10d supergravity, which live on the Dp-branes, can also be studied. In particular, one may associate the known structure of 4d theory with ${\cal N}=1$ vector multiplets with 10d theory with Dp-branes. The dictionary between string theory models and ${\cal N}=1$ 4d supergravity in such case will also involve the holomorphic vector coupling functions $f_{AB}(z)$.

In  \cite{Hamada:2018qef} an evidence was given that their proposal for the D7 brane action with 4-fermion terms with careful account of a compact volume of compactification leads to a 4d action with the vector multiplets.  

The action of ${\cal N}=1$ vector multiplets interacting with supergravity in case of the  holomorphic vector coupling $f_{AB}(z)$ is diagonal $f_{AB}(z) = f (z)\delta_{ AB}$
has the following relevant terms \cite{Freedman:2012zz}: 
\bea
&-{1\over 4}{\rm Re} f (z) \Big ( F_{\mu\nu}^A  F^{\mu\nu A } - 2 \bar \lambda \gamma^\mu D_\mu \lambda^A \Big ) +\cdots \cr
\cr
& + {1\over 4} e^{K(z, \bar z)\over 2}  f_{,\alpha} \overline \nabla^{\alpha} \overline W \, \bar \lambda^A P_L   \lambda^A  + h.c.\cr
\cr
&+{3\over 64} \Big [ {\rm Re} f (z) (\bar \lambda^A \gamma^\mu \gamma_* \lambda^A\Big ]^2\cr 
\cr
& -{1\over 16} f_{,\alpha}  \bar f_{,}{}^{\alpha} \bar \lambda^A P_L   \lambda^A \bar \lambda^B P_R   \lambda^B \ .
\label{FVP}\eea
This expression is in agreement with eqs. \rf{aux}, \rf{aux1} above.
Clearly,  4d supergravity if derived from 10d string theory with branes, should have no singularities and must have 4-fermion gaugino coupling.

If $ f (z)$ is a gauge coupling on N coincident D7 branes and 10d theory is compactified on a compact $T^2\times T^2\times T^2$ space with finite volume, one finds that indeed, the 4d action \rf{FVP} with terms bilinear and quartic in gaugino  follows from eq. \rf{AG},  as observed in \cite{Hamada:2018qef}.

\

In conclusion, here we have shown that the perfect square term in the 10d Einstein-Yang-Mills supergravity originates from the superspace geometry. Namely, the action includes a square of the first component of the supertorsion superfield, as shown in eq. \rf{T2}, which is a unique combination of the 3-form and a gaugino bilinear in eq. \rf{Y}. The same torsion superfield is present in gaugino equation of motion consistent with supersymmetry, as shown in \rf{EOM}. This means that a cubic gaugino coupling is present in gaugino equation of motion, and the quartic fermion coupling must be present on non-Abelian Dp-brane, as follows from the superspace geometry.

A better understanding of the impact of the 4-fermion terms  in condensing D7 branes, particularly with regard to a finite volume of compactification, would be useful for the purpose of  relating 10d string theory to cosmological observations described by the  4d physics.

\

\acknowledgments

I would like to thank  S. Ferrara, M. Dine, A. Linde,  E. McDonough, B. Nillsson and A. Tseytlin  for helpful comments and discussions. This work   is supported by SITP,  by the NSF Grant PHY-1720397, and by the Simons Foundation Origins of the Universe program (Modern Inflationary Cosmology collaboration).

\bibliography{lindekalloshrefs}
\bibliographystyle{utphys}

\end{document}